\documentclass[a4paper,11pt]{article}
\usepackage{jinstpub} 
\usepackage{lineno}
\usepackage{graphicx}
\usepackage{siunitx}
\usepackage{subcaption}
\usepackage{makecell} 
\usepackage{booktabs}

\title{Development of a Proton Therapy Research Beamline with FLASH and Minibeam Capabilities at the 18 MeV Bern Medical Cyclotron }

\author[1*]{Eva Kasanda\note[*]{Corresponding author.}}
\author[1]{Lars Eggimann}
\author[1]{Thierry Stammbach}
\author[1,2,3]{Pierluigi Casolaro}
\author[1,4]{Gaia Dellepiane}
\author[1]{Alexander Gottstein}
\author[5]{Jan Gruber}
\author[1]{Isidre Mateu}
\author[5]{Paolo Pellicioli}
\author[1,6]{Maria Vittoria Rossi}
\author[1,3]{Paola Scampoli}
\author[5]{Cristian Fernandez Palomo}
\author[1]{Saverio Braccini}

\affiliation[1]{Albert Einstein Center for Fundamental Physics, Laboratory for High Energy Physics, Universit\"{a}t Bern, Sidlerstrasse 5, 3012 Bern, Switzerland}
\affiliation[2]{INFN Sezione di Napoli, Complesso Universitario di Monte S. Angelo, 80126 Napoli, Italy}
\affiliation[3]{Department of Physics ‘‘Ettore Pancini’’, University of Napoli Federico II, Via Cintia, 80126 Napoli, Italy}
\affiliation[4]{Tera-Care Foundation, Chemin du Coq-d’Inde 8a, 1223 Cologny, Switzerland}
\affiliation[5]{Institute of Anatomy, Universit\"{a}t Bern, Baltzerstrasse 2, 3012 Bern, Switzerland}
\affiliation[6]{Department of Science and High Technology, Università degli Studi dell'Insubria, Via Valleggio 11, Como, Italy}

\emailAdd{eva.kasanda@unibe.ch}

\abstract{
Advanced radiotherapy approaches such as FLASH irradiation and spatially fractionated radiotherapy (SFRT) show potential to improve the therapeutic ratio, yet their biological mechanisms and optimal delivery parameters remain uncertain. Progress requires accessible proton research platforms with flexible temporal and spatial dose delivery. We report on the adaptation of the Beam Transfer Line (BTL) of the Bern Medical Cyclotron (BMC) for radiobiology research with FLASH and proton minibeam capabilities.
The BMC is optimized for the production of radionuclides for medical imaging, and is able to extract currents up to \SI{150}{\micro\ampere}. The \SI{18}{\mega\electronvolt} proton beam was passively shaped using collimators, scattering foils, and extended drift space to generate irradiation fields. A dosimetric framework was implemented using an in-beam ionization chamber and radiochromic film with LET-dependent corrections. Beam uniformity and SFRT profiles with various grid spacings were evaluated at realistic target distances.
The developed beamline enables stable delivery under controlled conditions in both conventional and FLASH regimes, spanning dose rates from 0.01 to \SI{100}{\gray\per\second}. Dose uniformity within a 20 mm radius was below 8\%. Film measurements confirmed the need for LET-dependent corrections and indicated that quantitative dosimetry in in-vitro setups is achievable with appropriate LET corrections. The low proton energy (\SI{15.54(12)}{\mega\electronvolt} extracted into air, \SI{8.14(28)}{\mega\electronvolt} delivered to cells in flask) facilitates compact SFRT implementation with well-resolved minibeams.
The adapted BMC provides a flexible and accessible platform for systematic pre-clinical proton radiobiology studies under varied dose-rate and spatial delivery conditions. This supports optimization of emerging modalities such as proton FLASH and SFRT and helps bridge accelerator technology and radiobiology.

}


\begin{document}
\maketitle
\flushbottom

\section{Introduction}


Emerging radiation therapy approaches such as ultra-high dose rate irradiation (FLASH) and Spatially Fractionated Radiation Therapy (SFRT) have demonstrated the potential for further improvements in normal tissue preservation and therapeutic efficacy \cite{Velalopoulou2021,Yang2025}, and can be extended to proton therapy. While promising pre-clinical and early clinical results have been reported, the biological mechanisms driving these effects remain poorly understood, and consensus on treatment planning parameters has yet to be established \cite{Yang2025, Lee2026}.

Progress in this field depends critically on accessible and flexible research platforms that allow systematic investigation of radiobiological responses under different delivery conditions. In particular, proton facilities capable of implementing both FLASH and SFRT in radiobiological settings are needed to clarify the underlying biology and guide clinical translation. Radiopharmaceutical production cyclotrons are well-suited for such studies, as they can easily extract proton currents high enough to deliver dose rates in the FLASH regime \cite{BarattoRoldn2018,BarattoRoldn2020,Constanzo2019,Ghithan2015, Fabbrizi2025}. 

The Bern Medical Cyclotron (BMC) laboratory, together with its beam transfer line, constitutes a suitable platform in this context. While the \SI{18}{\mega\electronvolt} cyclotron provides a high-current proton source, the beam transfer line enables a high degree of customization, allowing its use as an accessible proton therapy research facility with FLASH capabilities. 

In this work, we present a first beam shaping setup and dosimetry for this facility, with emphasis on its application to in-vivo and in-vitro proton therapy studies.

\section{Experimental Facility}
The BMC is an IBA Cyclone 18/18 HC proton cyclotron located at the Bern University Hospital (Inselspital), Switzerland \cite{Braccini2013}. 

A dedicated Beam Transfer Line (BTL) extends \SI{6.5}{\m} from the cyclotron to a research bunker that is fully independent from routine radiopharmaceutical operations, ensuring safe and flexible experimental access. The facility can deliver beam currents from a few pA up to 150 µA, covering the full range required for conventional dose rate irradiation and FLASH studies \cite{Auger2015}.

The extracted proton energy of 18.12(7) MeV \cite{Gottstein2026} is lower than that of typical clinical facilities, offering advantages for systematic radiobiology research such as irradiation with high-LET protons at shallow depths, reduced beam degradation, and simplified beam shaping. In particular, the latter enables straightforward implementation of SFRT using metallic collimators as thin as 0.5~mm.

\section{Experimental Setup}

\begin{figure}[h]
    \centering
    \begin{subfigure}[b]{0.7\textwidth}
           \includegraphics[width=\textwidth]{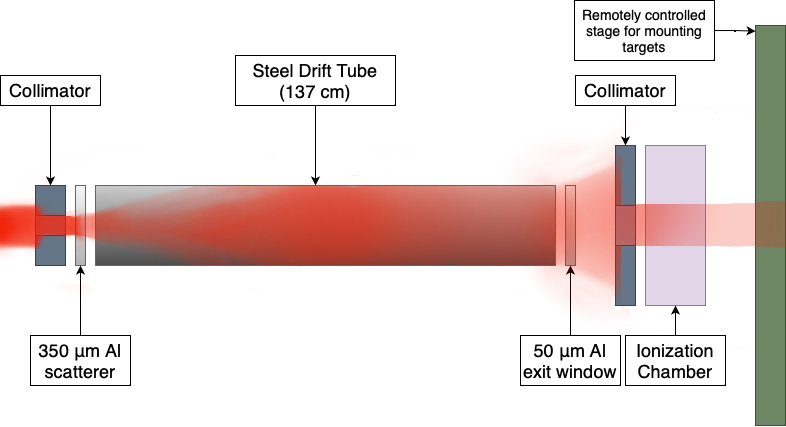}
        \caption{}
        \label{fig:subfig:cell-setup-sketch}
    \end{subfigure}
    \vfill
    \begin{subfigure}[b]{0.7\textwidth}
        \includegraphics[width=\textwidth]{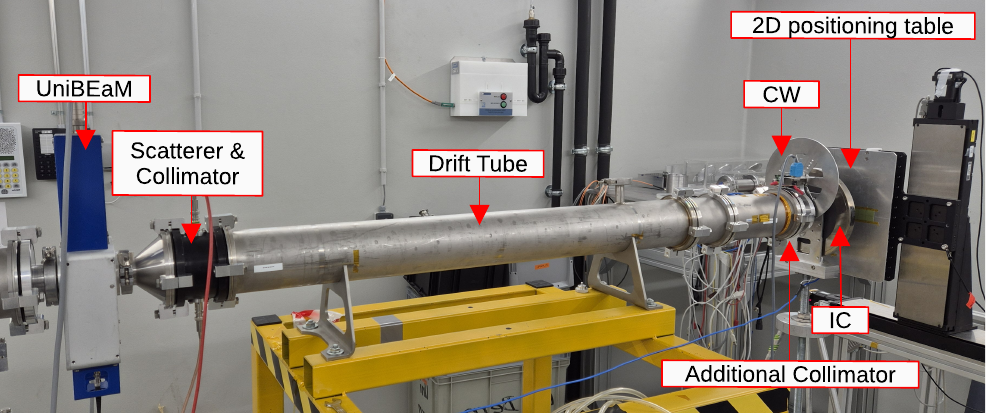}
        \caption{}
        \label{fig:subfig:cell-setup-photo}
    \end{subfigure}
    \caption{Experimental setup for beam shaping and current reduction in the BTL bunker. (a) Schematic of the beamline configuration. (b) Photograph of the setup with key components indicated.
   \label{fig:cell-setup}}
\end{figure}

Proton therapy studies at conventional rates require stable beam current over the course of several minutes. Therefore, the beam current delivered to the target is reduced to a dose rate in the conventional regime by collimating and scattering the beam within the BTL bunker. 
The experimental setup including beam shaping and diagnostic devices is shown in Figure~\ref{fig:cell-setup}.

\paragraph{Passive scattering setup. }\label{sec:passive-scattering}
The BTL is equipped with two sets of focusing quadrupole magnets, which are capable of focusing the beam down to a FWHM of a few \SI{}{\milli\meter}. For this application, minimal focusing is applied to extract a broad Gaussian at the position of the UniBEaM \cite{Potkins2017}, a two-dimensional beam profiler based on scintillating optical fibers. To reduce the proton flux  to the level required for the delivery of conventional proton therapy dose rates (0.01 -\SI{0.05}{\gray\per\second}), the defocused beam is centered at the UniBEaM, then passed through a pinhole collimator with a diameter of \SI{1}{\milli\meter}. For higher dose rates, a larger collimator aperture can be used (\SI{4}{\milli\meter}, \SI{10}{\milli\meter} or \SI{35}{\milli\meter}). 


In order to further increase the size and uniformity of the treatment field in this dose rate regime, a \SI{350}{\micro\meter}-thick aluminum scatterer is inserted immediately distal to the collimator. 
Downstream of the scatterer is a \SI{137}{\centi\meter}-long steel beam pipe, which acts as an extended drift space for the scattered beam to spread under vacuum into a wide Gaussian distribution. At the end of this drift tube, the beam is then extracted into air through a \SI{50}{\micro\meter}-thick aluminum vacuum window with a diameter of \SI{55}{\milli\meter}.  The length of the drift space is correlated with the reduction in the extracted proton flux. Therefore, the insertion of additional instrumentation between the scatterer and the exit window will result in a lower extracted dose rate. 

The mean beam energy at this extraction point has been experimentally determined to be \SI{15.54(12)}{MeV}, using a specific method developed by our group based on stacked-foil activation and a Bayesian statistical framework \cite{Gottstein2026}. The beam is then collimated once more in air down to the desired field size directly upstream of the in-beam monitoring chamber described in Section \ref{sec:dosimetry-setup}.

The established passive scattering setup has a dual purpose. First, it enables the reduction of the extracted dose rate from values exceeding \SI{200}{\gray\per\second} down to a range relevant for conventional biological experiments (on the order of \SI{0.01}{\gray\per\second}). Second, it improves the uniformity of the extracted beam and increases the size of the treatment field plateau beyond the limits of what can be achieved with magnetic defocusing of the beam. Additionally, the shape of the scattered beam is much less susceptible to fluctuations over the course of the irradiation. 

\begin{figure}[h]
    \centering
    \includegraphics[width=0.7\textwidth]{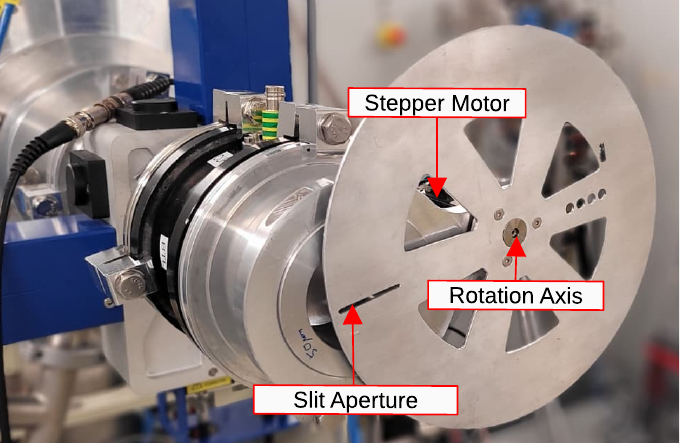}
    \caption{Custom-built chopper wheel for precise dose delivery at high dose rates.
    \label{fig:chopperwheel}}
 \end{figure}

\paragraph{Beam chopper. } For FLASH dose rates, precise control of short irradiation windows is required. To achieve this, a remotely controlled beam Chopper Wheel (CW) has been developed and is shown in Figure~\ref{fig:chopperwheel}. The CW consists of a circuit board, a stepper motor (AnalogDevices PD28-1021~\cite{PD1021DatasheetProduct}, controlled using the PyTrinamic Python library~\cite{PytrinamicTRINAMICsPython}) and an aluminum wheel with a slit aperture. The wheel is mounted on the stepper motor and rotates to achieve the desired beam pulse. Two different aperture widths are available for the wheel, enabling controlled dose delivery over the full range of dose rates that can be extracted from the cyclotron (0.01 Gy/s - 2200 Gy/s). The whole construction is mounted on the beam pipe such that the slit passes through the beam as the wheel rotates, as shown in Figure \ref{fig:chopperwheel}. The system operates at rotation frequencies between 0.2 and \SI{4}{\Hz}, corresponding to pulse durations from \SI{0.5}{\second} down to approximately \SI{1}{\milli\second}, depending on slit width. This allows the modulation of both the temporal structure and delivered dose per pulse. 

\begin{figure}[h]
	\centering
	\begin{subfigure}[t]{0.43\textwidth}
		\centering
		\includegraphics[width=0.9\textwidth]{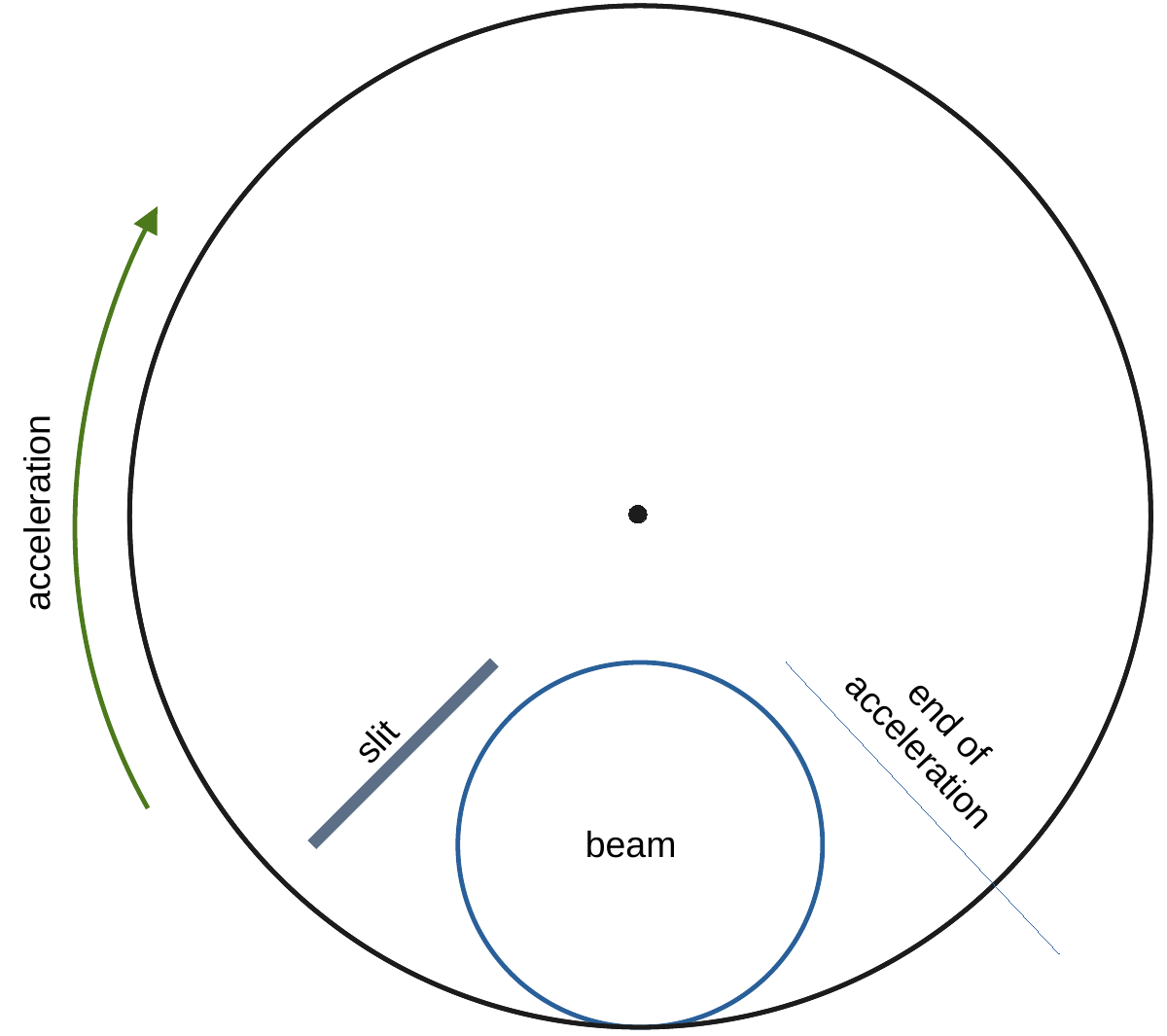}
		\caption{}
		\label{fig:cw_scetch}
	\end{subfigure}
    \hfill
	\begin{subfigure}[t]{0.53\textwidth}
		\centering
		\includegraphics[width=\textwidth]{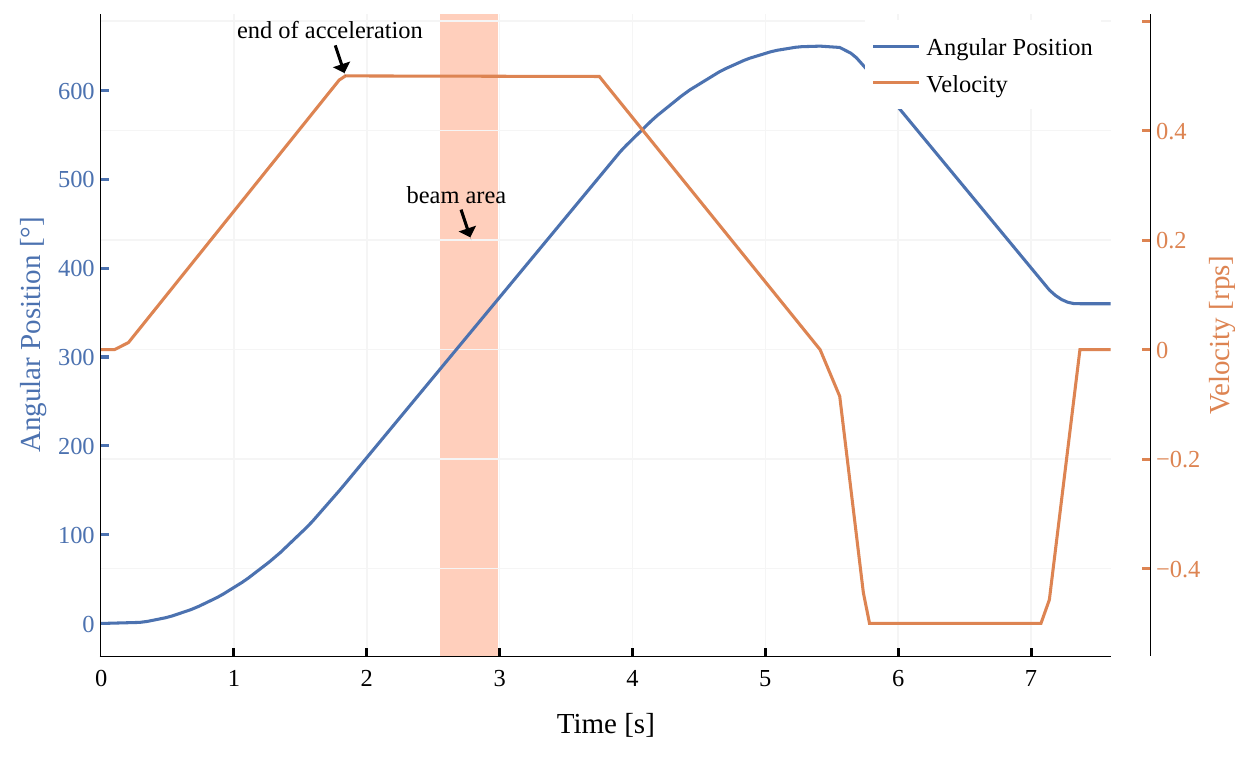}
		\caption{}
		\label{fig:cw_test_path}
	\end{subfigure}
	\caption{Chopper wheel rotation profile. (a) Schematic illustrating the required motion of the slit relative to the beam area. (b) Example of angular position and velocity for a 0.5 Hz rotation setting.}
	\label{fig:cw_rotation}
\end{figure}

Uniform dose delivery requires constant angular velocity during beam transmission through the slit. Therefore, the CW follows a distinct rotation profile, such that the acceleration and deceleration phases are outside of the beam area. Figure~\ref{fig:cw_scetch} illustrates in which direction the acceleration starts. The CW must reach the desired velocity before reaching the ``end of acceleration'' line. From there, the velocity stays constant and the deceleration only begins after the slit leaves the beam area. Figure~\ref{fig:cw_test_path} shows the profile of a stable pulse rotation pattern. 

 \paragraph{Target mounting. } At the end of the beamline, a remotely controlled two-axis positioning stage is installed. It is controlled by stepper motors that enable adjustment in increments of \SI{100}{\micro\meter}. This stage is aligned with the center of the beam with the aid of ceiling- and wall-mounted lasers in the BTL bunker to a precision of approximately $\pm$ \SI{1}{\milli\meter}. The stage is outfitted with an anodized aluminum mounting plate and a beam dump.

\paragraph{SFRT collimation. }
The low energy of the protons extracted from the beamline greatly facilitates the collimation of the beam for SFRT purposes, as the short proton range allows thin materials to fully stop the beam. For proton energies in the <\SI{20}{\mega\electronvolt} range, thicknesses of 0.5 mm of tungsten or 1.7 mm of aluminum are sufficient to completely stop the beam, enabling the use of compact tungsten collimators which are thin enough to be made with wire electrical discharge machining (WEDM). With WEDM, apertures cut to 200 \textmu m wide on a 1 mm thick tungsten plate are possible with about 5 \textmu m precision. The same collimator can be used with orthovoltage x-rays as well, facilitating comparative studies between proton and photon irradiation conditions. In this study, we investigate the impact of proton scattering on the dose distribution and the lateral penumbras of the SFRT beamlets.

\paragraph{Dosimetry. }\label{sec:dosimetry-setup}
The proton beam is monitored by an in-beam ionization chamber (PTW Freiburg GmbH, Monitor Chamber Type 786), \cite{PTW}, which provides a real-time signal proportional to the total proton current. This ionization chamber (IC) has been calibrated to the dose rate delivered to the target using EBT-3 Gafchromic films \cite{Ashland2023}, and using a custom-built insertable Faraday Cup (FC) as a reference for absolute fluence. At the low proton energies used in this setup (approximately \SIrange[]{8}{15}{\mega\electronvolt} at the target), the beam operates near the Bragg peak, where LET is high and rapidly varying. LET-dependent quenching effects in the radiochromic films are corrected using experimentally derived factors (Section \ref{sec:gaf-methods}), which introduce significant uncertainties (5.5\%) into the dose calculation and are, at the moment, a limiting factor in the precision of the dosimetry. 

\section{Methods and Measurements}

\subsection{Radiochromic film calibration}
Radiochromic films were selected for their high spatial resolution, which is required for SFRT characterization, despite their known LET-dependent response. The films are scanned using an Epson Perfection V850 Pro flatbed scanner with a resolution of \num{400} dpi. Doses were evaluated using the weighted grey pixel value \cite{BT601StudioEncoding} and a calibrated Green-Saunders equation as a fitting function \cite{campajolaAbsoluteDoseCalibration2017}. The calibration was performed in the range of \SIrange[]{0.1}{15}{\gray} with a $^{60}$Co teletherapy unit gamma source at the Swiss National Institute for Metrology (METAS) \cite{stucki_muench_quintel_2003}. 

\subsection{Dose rate calibration}\label{sec:dose-rate-calibration-methodes}

The extracted proton beam current depends on multiple cyclotron control parameters and is not reliably reproducible, as it is influenced by factors such as operating temperature and variations in the geometry of beam extraction components between maintenance cycles. To enable comparison of dose-rate measurements across different collimator apertures, a stable reference quantity representing the upstream proton flux is required. This quantity is measured using an insertable FC positioned directly upstream of the collimator.

Direct correlation between FC current and absorbed dose at the irradiation position is not possible, as radiochromic films provide only an integrated dose measurement and cannot be read out in real time. Therefore, an in-beam ionization chamber (IC) is used as an intermediate detector to link beam current to delivered dose.

A calibration of the dose rates achieved with the passive scattering setup was performed for four collimator apertures (see Section~\ref{sec:passive-scattering}). The procedure consists of two steps: (i) calibration of the IC response to absorbed dose, and (ii) calibration of the IC current to the FC current upstream of the collimator.

\medskip
\noindent\textbf{\textit{Calibration of ionization chamber to EBT-3 film}}\\
The IC was first calibrated against radiochromic film measurements to establish the relationship between collected charge and absorbed dose at the irradiation position. This was achieved by correlating the integrated IC charge during a beam pulse with the dose recorded on film, yielding a calibration factor that allows conversion of IC current into dose rate for a given experimental configuration.

For these measurements, the beam was collimated to a \SI{10}{\milli\meter} $\times$ \SI{10}{\milli\meter} field immediately upstream of the IC to ensure a well-defined field size. This is required because the IC signal is proportional to the total incident proton current, whereas the dose rate depends on proton fluence.

Due to beam divergence in air and the energy dependence of the IC response, the calibration is sensitive to the geometry of beamline elements downstream of the exit window. In particular, the distance between the IC and the irradiation target affects the dose-rate-to-current conversion through (i) fluence reduction caused by beam divergence and (ii) changes in proton energy and LET due to energy loss in air. The relevant distances for the present setup are summarized in Table~\ref{tab:distances}. Consequently, the IC calibration is specific to the given experimental configuration.

\begin{table}[ht]
\centering
\caption{Relevant distances in the experimental setup.}
\begin{tabular}{lc}
\hline
\textbf{Component} & \textbf{Distance (\si{\centi\meter})} \\
\hline
Exit window to $1\,\text{cm} \times 1\,\text{cm}$ collimator & 0.6 \\
Collimator to ionization chamber (IC) & 0.6 \\
Ionization chamber (IC) to target position & 4.1 \\
\hline
\end{tabular}
\label{tab:distances}
\end{table}

Measurements were performed for circular collimator apertures of 1, 4, 10, and \SI{35}{\milli\meter} diameter, each combined with a \SI{350}{\micro\meter}-thick aluminum scatterer positioned downstream of the collimator (Figure~\ref{fig:subfig:cell-setup-sketch}). The arc current of the ion source was varied between 1~mA and 29~mA to span the accessible dose-rate range, corresponding to extracted proton currents of approximately \SIrange[]{10}{300}{\nano\ampere} incident on the collimator.

\medskip
\noindent\textbf{\textit{Calibration of Faraday cup to ionization chamber}}\\
In a second step, the FC was used to establish a relationship between the upstream proton flux and the IC current for each collimator configuration. The FC provides an absolute measurement of the extracted proton current upstream of the collimator and scatterer, while the IC measures the transmitted beam in air at the end of the beamline. 

By correlating FC and IC signals over the same range of cyclotron ion source currents, a linear relationship between upstream proton flux and IC current is obtained. Combined with the IC-to-dose calibration, this enables determination of the dose rate at the irradiation position as a function of extracted beam current for each collimator aperture. The FC measurement serves as an absolute reference, as it directly measures deposited charge and is independent of LET effects.

For each collimator aperture, measurements were performed for cyclotron ion source currents up to 29~mA. The beam was collimated to a diameter of \SI{62}{\milli\meter} upstream of the FC and to a \SI{40}{\milli\meter} $\times$ \SI{40}{\milli\meter} field upstream of the IC.

\subsection{Radiochromic film LET correction}\label{sec:gaf-methods}
Gafchromic EBT-3 films were used for dose measurements; however, their response to low-energy protons requires correction due to LET-dependent effects.

For proton energies below approximately 20~MeV, EBT-3 Gafchromic films exhibit a non-linear dose response (quenching) due to the high linear energy transfer (LET) of the incident particles, resulting in an underestimation of the absorbed dose. This is quantified as the energy-dependent relative efficiency of the film, $RE$. In addition, protons lose energy in the polyimide lamination layers of the film, increasing the effective LET within the active layer relative to that at the film surface. This effect is quantified by a lamination layer correction factor  L, defined as the ratio of LET in the active layer to that at the film surface. This effect causes an overestimation of the dose, competing with the quenching effect.
The corrected dose can then be calculated as follows.
\begin{equation}
	D = D_{\text{meas}} \cdot \frac{L}{RE}
\end{equation}
The determination of these correction factors is described below.

\begin{figure}[h]
	\centering
	\begin{subfigure}[t]{0.40\textwidth}
		\centering
        \includegraphics[width=\textwidth]{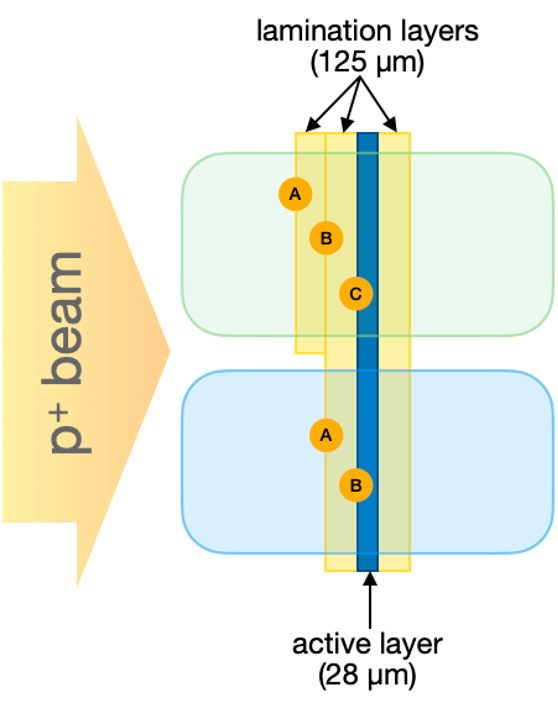}
        \caption{}
        \label{fig:lamination_energies}
	\end{subfigure}
    \hfill
	\begin{subfigure}[t]{0.48\textwidth}
		\centering
        \includegraphics[width=\textwidth]{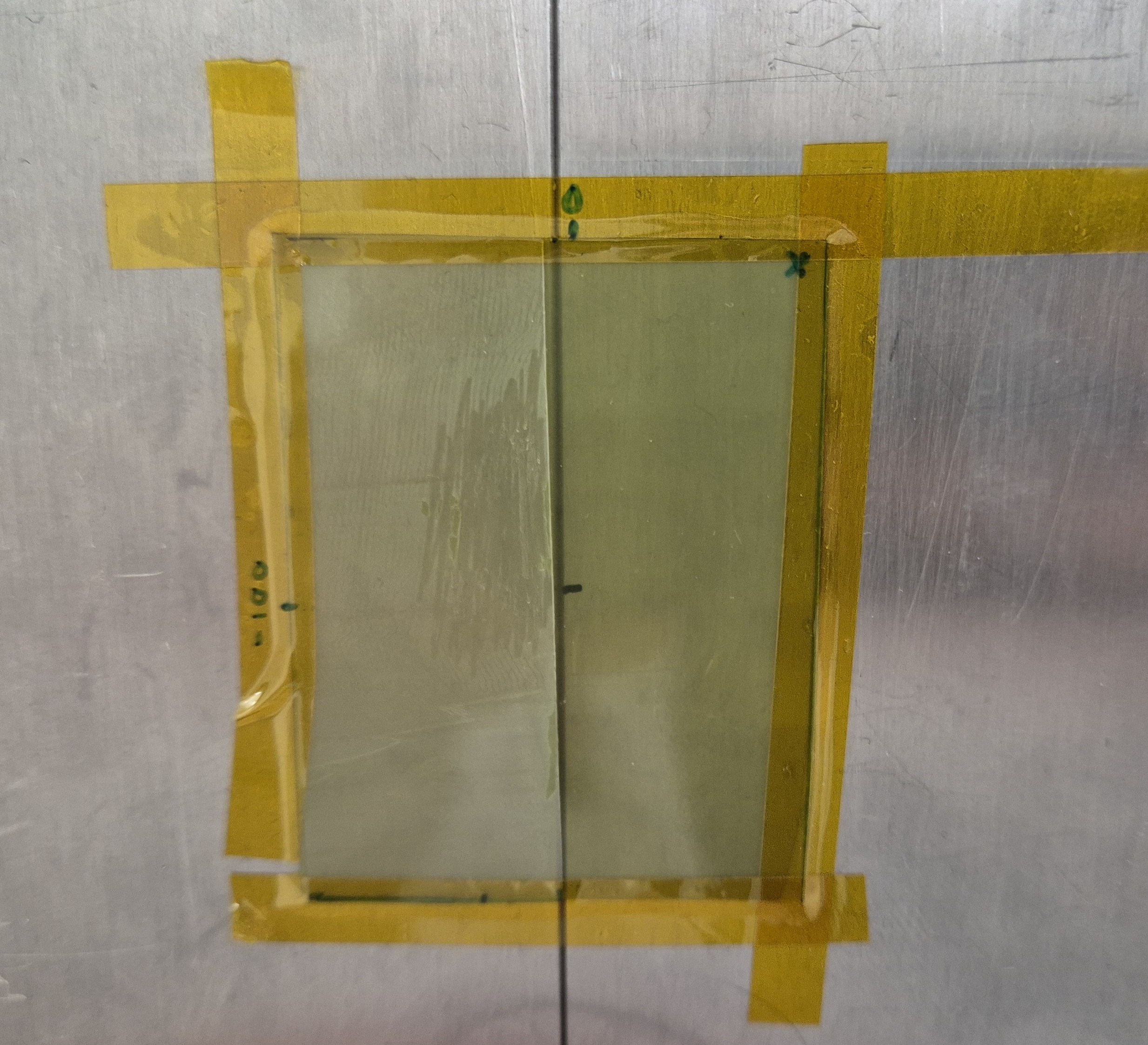}
        \caption{}
        \label{fig:lamination_layer_photo}
	\end{subfigure}
	\caption{Measurement of the lamination layer correction factor. (a) Schematic of the EBT-3 film with an additional upstream lamination layer partially covering the irradiation field; markers A, B, and C indicate positions used for energy and LET evaluation. (b) Photograph of the prepared film with partial lamination coverage.}
	\label{fig:lamination}
\end{figure}

\noindent\textbf{\textit{Lamination Layer Correction Factor $L$}}\\
The effect of the energy loss in the lamination layer of the EBT-3 film on the measured dose can be measured directly by partially covering it with an additional \SI{125}{\micro\meter}-thick polyimide lamination layer, as shown in Figure \ref{fig:lamination_energies} and irradiating it at the sample position to ensure identical proton energy and fluence, including the effects of beam divergence and scattering. For this study, the film was placed inside a standard cell culture flask to reproduce the geometry of in-vitro irradiation experiments in which the proton beam traverses the flask wall before reaching the cell layer. From the dose ratio between the two sections of the irradiated film, the relative dose delivered to the cell layer (as opposed to the film’s active layer) can be inferred.

To estimate the proton energy at each marked position in Figure \ref{fig:lamination_energies}, LISE++ simulations of the beamline geometry were performed. The simulations assumed an initial beam energy of \SI{18.12}{\mega\electronvolt} and included all relevant components listed in Table \ref{tab:beamline_layers}.


\begin{table}[h!]
\centering
\vspace{0.5\baselineskip}
\caption{Beamline geometry layers used for LISE++ simulation of extracted beam energy}
\begin{tabular}{lll}
\toprule
\textbf{Item} & \textbf{Material} & \textbf{Thickness / Length} \\
\midrule
Scatterer & Aluminum & \SI{350}{\micro\meter} \\
Beam pipe & Vacuum & \SI{136.9}{\centi\meter} \\
Exit window & Aluminum & \SI{50}{\micro\meter} \\
Air gap & Air & \SI{3}{\centi\meter} \\

\addlinespace[0.5em]   

PTW ionization chamber & 
\makecell[l]{Graphite \\ Aluminum \\ Kapton \\ Air} & 
\makecell[l]{\SI{180}{\micro\meter} \\ \SI{10}{\micro\meter} \\ $3 \times \SI{25}{\micro\meter}$ \\ $2 \times \SI{2.5}{\milli\meter}$} \\

\addlinespace[0.5em]   

Air gap & Air & \SI{2.5}{\centi\meter} \\
Cell flask wall & Lucite & \SI{1.5}{\milli\meter} \\
\bottomrule
\end{tabular}
\label{tab:beamline_layers}
\end{table}

The extracted proton energy at the distal edge of the cell flask wall is 8.14(28)~MeV, in agreement with monitor reaction energy measurements previously reported for this beamline \cite{Gottstein2026}. The proton energy at the irradiation position is highly sensitive to the spacing of beamline elements distal to the exit window, due to energy loss in air. As these distances are not precisely reproducible, configuration-specific calibration is required for each experimental setup.

The LISE++ simulations were used to determine the proton energy distribution after 0, 1, and 2 additional layers of \SI{125}{\micro\meter}-thick polyimide downstream of the flask wall (corresponding to the positions A, B, and C in Figure \ref{fig:lamination_energies}). From these distributions, the dose-weighted LET in water was calculated for each marker position, summarized in Table \ref{tab:lamination_energies}.  While the dose-weighted LET accounts for the energy distribution of the proton beam, additional uncertainties arising from the simulation and material properties are not explicitly quantified in this work and should be considered when interpreting the corrected dose values.

\begin{table}[h]
\centering
\vspace{0.5\baselineskip}
\caption{Beam energy and LET values at relevant depths in the irradiation setup according to LISE++ \cite{LISERareIsotope} calculations using the material configuration outlined in Table \ref{tab:beamline_layers}. Uncertainties in energy correspond to the standard deviation of the energy distribution calculated by LISE++. }\label{tab:lamination_energies}
\begin{tabular}{c S[table-format=1.2] S[table-format=1.4]}
\hline
Marker & {Beam Energy (MeV)} & {LET (keV/$\mu$m)} \\
\hline
A & 8.14(28) & 5.39 \\
B & 7.19(31) & 5.96 \\
C & 6.13(36) & 6.77 \\
\hline
\end{tabular}
\end{table}

Since the dose delivered to a material is proportional to the LET of the incident particle, the LET ratio between two configurations is equivalent to the corresponding dose ratio. The simulated LET ratio is 1.11 between markers A and B and 1.13 between markers B and C. Over this small energy range, the proton stopping power of water can be approximated as linear with a relative error below 2\% \cite{LISERareIsotope, Ziegler2010}. This makes the calculation relatively insensitive to small uncertainties in the simulated extracted energy. Additionally, because both halves of the film are irradiated simultaneously, the measurement is independent of the integrated dose, as the correction factor only depends on the ratio of the doses delivered to each portion of the film.

The dose ratio between markers A and B represents the correction needed to estimate the dose delivered to a thin cell layer on the distal face of the flask wall. This correction factor can be inferred experimentally, and is approximated by measuring the dose ratio between the two sections of the irradiated film (green and blue regions in Figure \ref{fig:lamination_energies}), which corresponds to the LET ratio between markers B and C.

This measurement quantifies the impact of the energy loss within the lamination layer on the measured dose, but does not account for the relative efficiency of the films for protons in this configuration in comparison to the gamma rays with which the films were calibrated.

\noindent\textbf{\textit{Relative Efficiency Correction Factor $RE$}}:\\
As described in \cite{SanchezParcerisa2021}, Relative Efficiency (RE) refers to the ratio of dose required to induce a certain film darkening from reference radiation (typically photon radiation) compared to high Linear Energy Transfer (LET) radiation (proton radiation). It is a measure used to quantify how effectively a film responds to different types of radiation.
For high energy protons, the of Gafchromic EBT-3 films generally exhibit a response largely independent of radiation type, which is one of their key features. However, when approaching the Bragg Peak region (lower energy protons), where the proton LET increases dramatically, the films have been shown to underrespond significantly \cite{Reinhardt2015}. This underresponse is attributed to the local saturation of polymerization sites within the film. When the LET is high, a large amount of energy is deposited in a localized region, activating all available polymerization sites along an ion track. Once all the sites are saturated, additional deposited energy cannot contribute to further polymerization, leading to a reduced color change and underestimation of the dose. This effect is often referred to as LET quenching.

The RE factor can be calculated using the following equation:
\begin{equation}\label{eq:re}
	RE(LET) = 1 - A \cdot LET^B
\end{equation}
where $A = \num{0.0117 \pm 0.0016}$ \SI{}{\micro\meter\per\kilo\electronvolt} and $B = \num{1.01 \pm 0.04}$ are calibration parameters taken from \cite{SanchezParcerisa2021}. The LET in \SI{}{\kilo\electronvolt\per\micro\meter} is calculated using the following equation:
\begin{equation}\label{eq:let}
	LET = a \cdot e^{b E_s} + c \cdot e^{d E_s}
\end{equation}
where $a = \SI{4.1(15)e5}{\kilo\eV\per\micro\m}$, $b = \SI{2.88 \pm 0.12}{\per\mega\eV}$, $c = \SI{22.5 \pm 1.9}{\kilo\eV\per\micro\m}$ and $d = \SI{0.142 \pm 0.013}{\per\mega\eV}$ are calibration parameters taken from \cite{SanchezParcerisa2021}. As one can see the uncertainties of these parameters are quite large and constitute a significant contribution to the overall dose uncertainty. This correction factor also needs to be determined for each setup since it depends on the proton energy spectrum at the point of irradiation.

\subsection{Spatially fractionated radiation therapy}
\begin{figure}[h]
 \centering
    \includegraphics[width=0.4\textwidth]{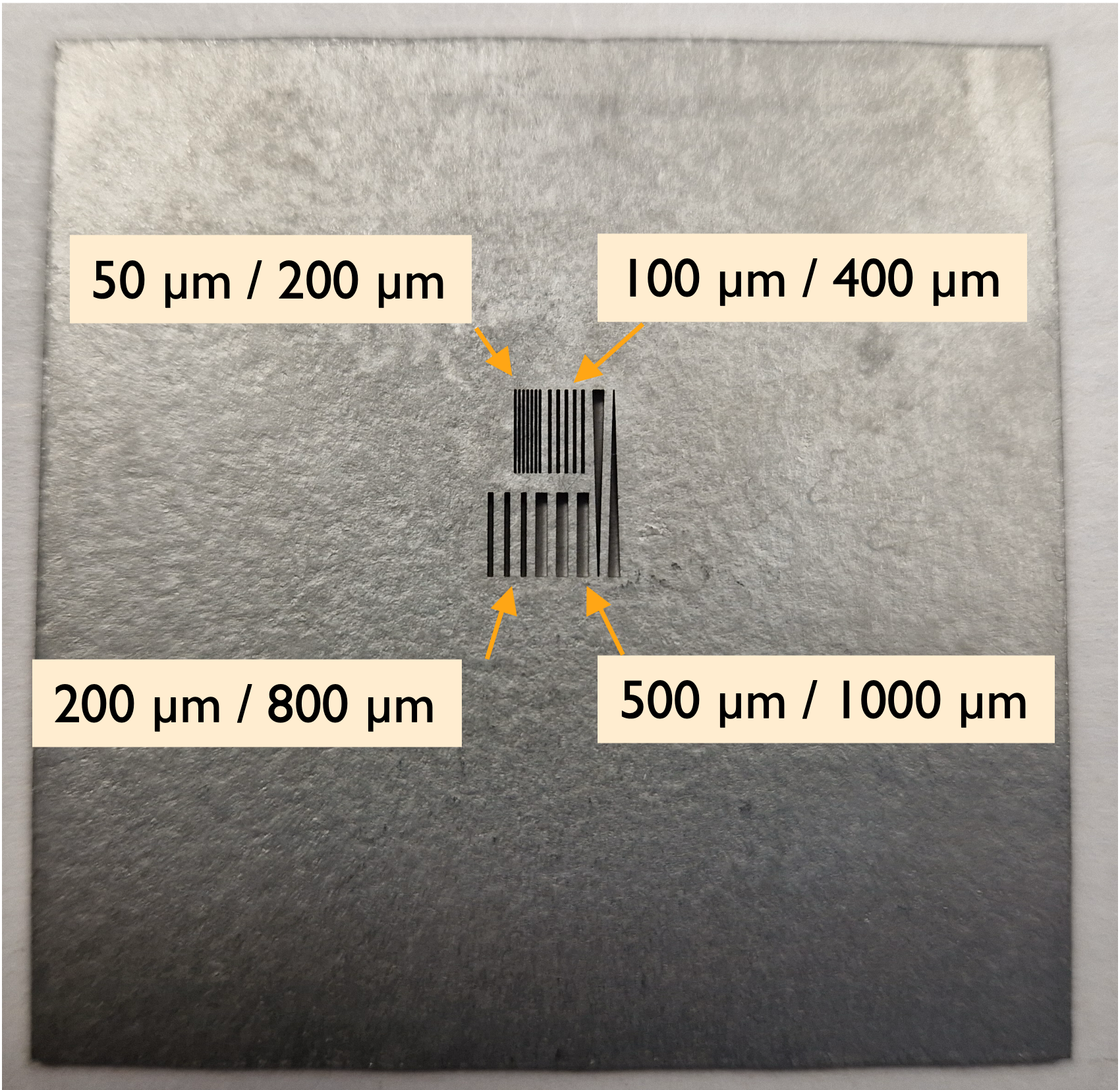}
    \caption{Collimator used to evaluate SFRT spatial resolution at target position for various aperture configurations. Labels indicate the aperture width and centre-to-centre distance for each configuration.
    \label{fig:sfrt-coll}}
 \end{figure}

Due to the passively scattered nature of the beam and its relatively high LET, beam divergence in air over a range of a few millimetres can have a significant impact on complex dose profiles such as those delivered in an SFRT treatment field. To evaluate the impact of the spacing between the SFRT collimator and the target on the delivered field, an EBT-3 Gafchromic film was placed at the target position downstream of a test collimator at distances of 0, 2, and \SI{4}{\milli\meter} in air from the distal surface of the collimator. The test collimator, shown in Figure \ref{fig:sfrt-coll}, consists of four regions with different aperture spacings, and a fifth section containing a smoothly varying wedge-shaped aperture. 

\section{Results}
\subsection{Beam Uniformity}\label{sec:uniformity-results}
\begin{figure}[h]
 \centering
    \includegraphics[width=1.0\textwidth]{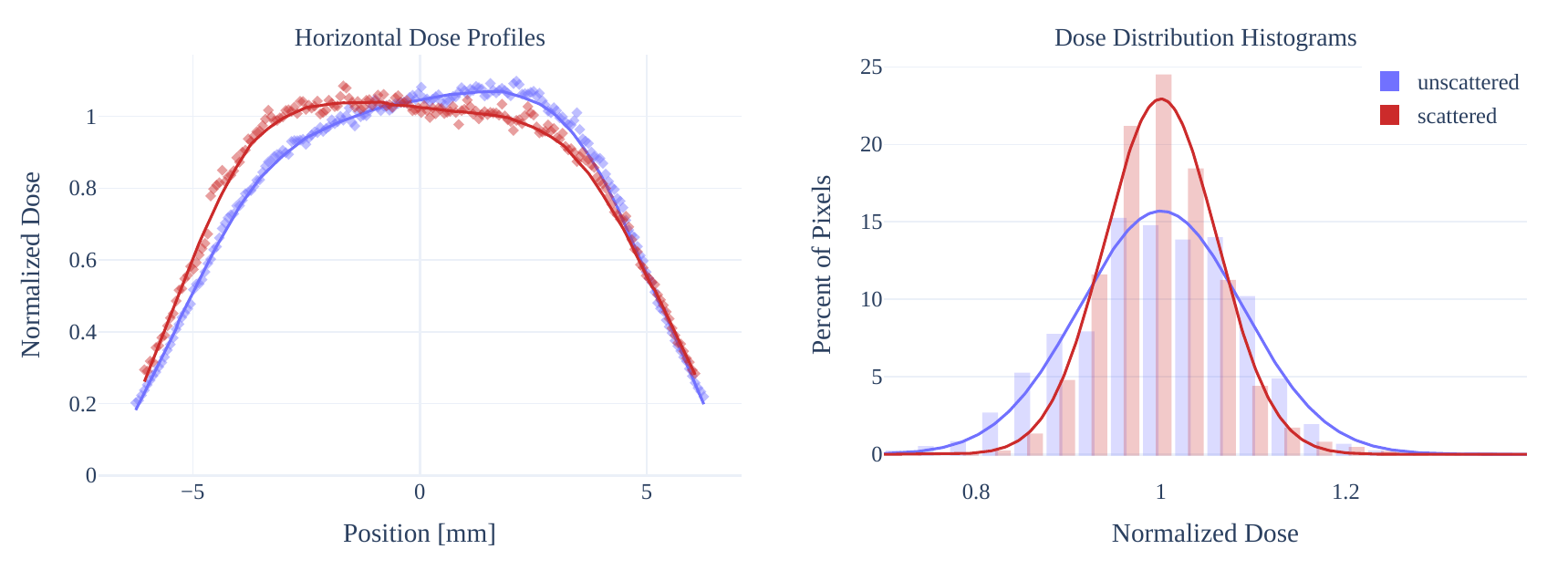}
    \caption{(left) Horizontal dose profiles for scattered and unscattered configurations obtained with a \SI{10}{\milli\meter} $\times$ \SI{10}{\milli\meter} square collimator. (right) Corresponding dose distributions within a \SI{6}{\milli\meter} $\times$ \SI{6}{\milli\meter} region centered on the beam.
    \label{fig:beamprofile-short-vs-long}}
 \end{figure}

Figure \ref{fig:beamprofile-short-vs-long} compares the beam uniformity obtained for a \SI{1}{\centi\meter^2} collimated field using magnetic defocusing and passive scattering. In this configuration, the field size is limited by the use of the chopper wheel (Figure \ref{fig:chopperwheel}), which enables delivery of a controlled dose of below \SI{10}{\gray} at the dose rates of the unscattered beam ($\approx$ 150~Gy/s), allowing the subsequent evaluation of the extracted dose profile on EBT-3 film. The left panel shows the horizontal dose profile (averaged over a vertical range of 20 pixels) obtained with and without the use of the scatterer. 

The standard deviation of the dose distribution within a \SI{6}{\milli\meter} $\times$ \SI{6}{\milli\meter} region centred on the beam is 8.84\% for the unscattered beam and 6.03\% for the scattered beam, as shown by the histograms in the right panel. The unscattered beam profile exhibits a pronounced asymmetry, which is prone to observable short-term drifts under the investigated operating conditions, attributed to temperature fluctuations in the cyclotron and variations in proton extraction conditions. In contrast, the scattered beam profile is more stable against these fluctuations in shape, with residual asymmetry primarily arising from alignment uncertainties.

\begin{figure}[h]
 \centering
    \includegraphics[width=1.0\textwidth]{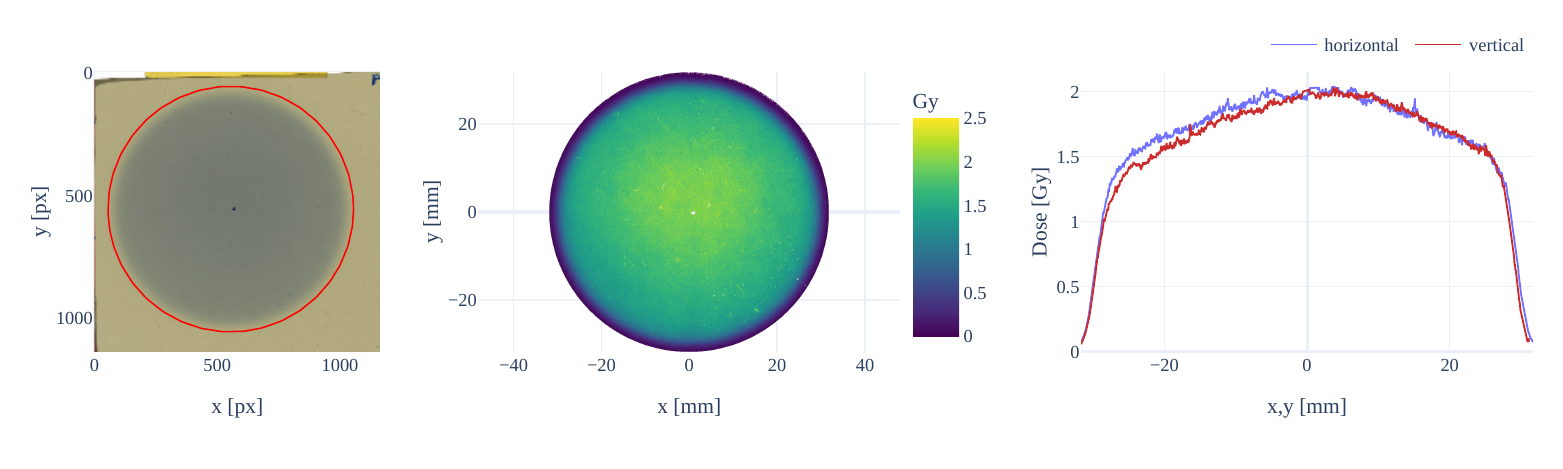}
    \caption{Dose distribution over the full extracted field (55 mm-diameter exit window). Left: scanned film. Centre: reconstructed dose map. Right: horizontal and vertical dose profiles illustrating beam uniformity.
    \label{fig:full-field-1mm}}
 \end{figure}

Figure \ref{fig:full-field-1mm} shows the dose profile obtained using the full extracted field, limited by the 55 mm-diameter circular exit window. 

 The dose profile was measured using the 1 mm collimator aperture, without the use of the chopper wheel. The left panel shows a scan of the irradiated film. The central panel shows the corresponding dose. The right panel shows the horizontal and vertical dose profile, illustrating the uniformity of the beam over the full field. Within a \SI{20}{\milli\meter} radius from the beam centre, the standard deviation of the delivered dose is 7.96\%.

When a cell flask is positioned in the treatment field, the standard deviation of the dose delivered to a \SI{30}{\milli\meter} $\times$ \SI{35}{\milli\meter} area is 7.70\%.

\subsection{Dose Rate Calibration}

\begin{figure}[hb]
\centering
\includegraphics[width=1.0\textwidth]{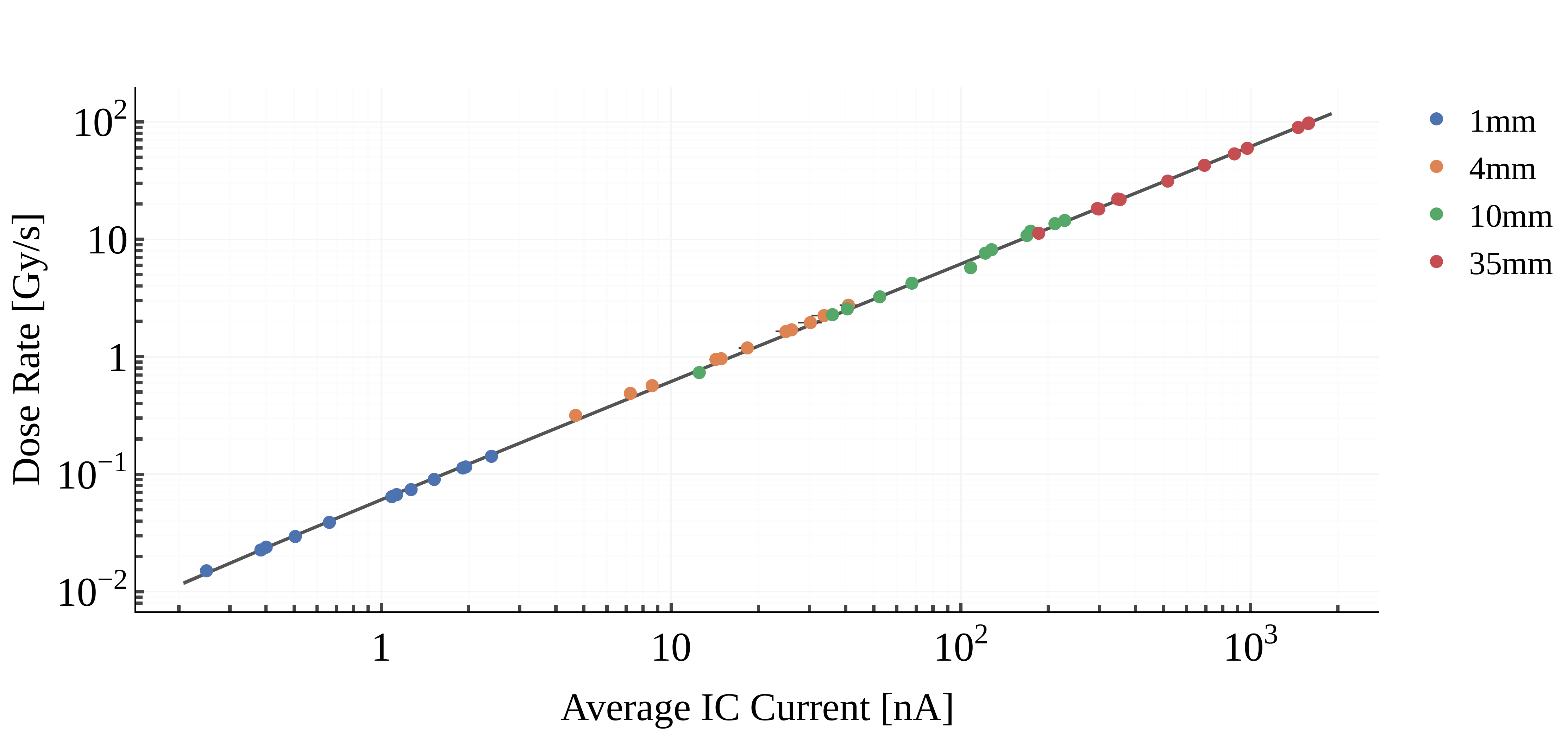}
\caption{Dose rate (obtained from film scans) at the target position as a function of measured IC current, for collimator diameters of 1, 4, 10, and \SI{35}{\milli\meter}. Error bars are included but not visible. Line represents linear fit of entire data set.}
\label{fig:dose-rate-range-IC}
\end{figure}

The calibration of the IC response to dose rate for all collimator apertures is summarized in Figure~\ref{fig:dose-rate-range-IC}. A linear relationship between IC current and dose rate is observed over the full investigated range.

Small variations in the calibration slope between collimator apertures are observed, which can be attributed to slight differences in detector positioning after collimator exchange. Nevertheless, the combined data set is well described ($R^2 = 0.9998$) by a single linear function:
\begin{equation}\label{eq:ICcal}
    DR = \left(6.17(6) \times 10^{7}\,\frac{\mathrm{Gy}}{\mathrm{s\cdot A}}\right)I_{IC} 
    - 0.0010(5)\,\frac{\mathrm{Gy}}{\mathrm{s}}.
\end{equation}

This calibration does not exhibit any significant saturation effects in the investigated range of dose rates, and is used moving forward to relate the extracted dose rate to other relevant diagnostic quantities without needing to continuously rely on film analysis. 

The next calibration relates the extracted dose rate to the absolute beam current measured by the FC. The relationship between FC and IC currents for each collimator aperture is given in Table~\ref{tab:I_FC(I_IC)}. A linear dependence is observed in all cases, linking the upstream proton current to the transmitted beam measured at the IC.

\begin{table}[h]
\centering
\renewcommand{\arraystretch}{1.3}
\begin{tabular}{ccc}
\hline
Collimator aperture & \multicolumn{2}{c}{$I_{IC}(I_{FC}) = a I_{FC} + b$} \\
                   & a & b [\SI{}{\ampere}]\\
\hline
1 mm  & $\SI{0.114539(35)}{}$ & $\SI{1.024(28)e-10}{}$ \\
4 mm  & $\SI{2.3392(9)}{}$ & $\SI{-1.2(6)e-10}{}$ \\
10 mm & $\SI{11.199(5)}{}$ & $\SI{-1.484(20)e-08}{}$ \\
35 mm & $\SI{84.428(35)}{}$ & $\SI{8.75(35)e-08}{}$ \\
\hline
\end{tabular}
\caption{IC current as a function of FC current for different collimator apertures.}
\label{tab:I_FC(I_IC)}
\end{table}

Combining the IC dose calibration with the FC–IC current relationship yields a direct conversion from FC current to dose rate at the target for each collimator aperture. A scaling factor of 14.247 was applied to account for the different IC collimation conditions used in the two calibration steps, including the change in exposed detector area and the field-size-dependent impact of beam non-uniformities.

The resulting linear fit parameters for the dose rate as a function of FC current are summarized in Table~\ref{tab:DR(I_FC)}. The intercepts are consistent with zero within uncertainties, indicating no significant systematic offset and supporting the assumption of linear detector response.

\begin{table}[h]
\centering
\renewcommand{\arraystretch}{1.3}
\begin{tabular}{ccc}
\hline
Collimator aperture & \multicolumn{2}{c}{$\mathrm{DR}(I_{FC}) = a I_{FC} + b$} \\
                   & a [\SI{}{\gray\per\second\per\ampere}] & b [\SI{}{\gray\per\second}]\\
\hline
1 mm  & $\SI{4.954(32)e+05}{}$ & $\SI{-0.00030(32)}{}$ \\
4 mm  & $\SI{1.019(4)e+07}{}$   & $\SI{-0.0019(23)}{}$ \\
10 mm & $\SI{4.859(19)e+07}{}$  & $\SI{-0.077(9)}{}$ \\
35 mm & $\SI{3.688(15)e+08}{}$  & $\SI{0.16(7)}{}$ \\
\hline
\end{tabular}
\caption{Dose rate as a function of Faraday cup (FC) current for different collimator apertures.}
\label{tab:DR(I_FC)}
\end{table}

\begin{figure}[h]
\centering
\includegraphics[width=1.0\textwidth]{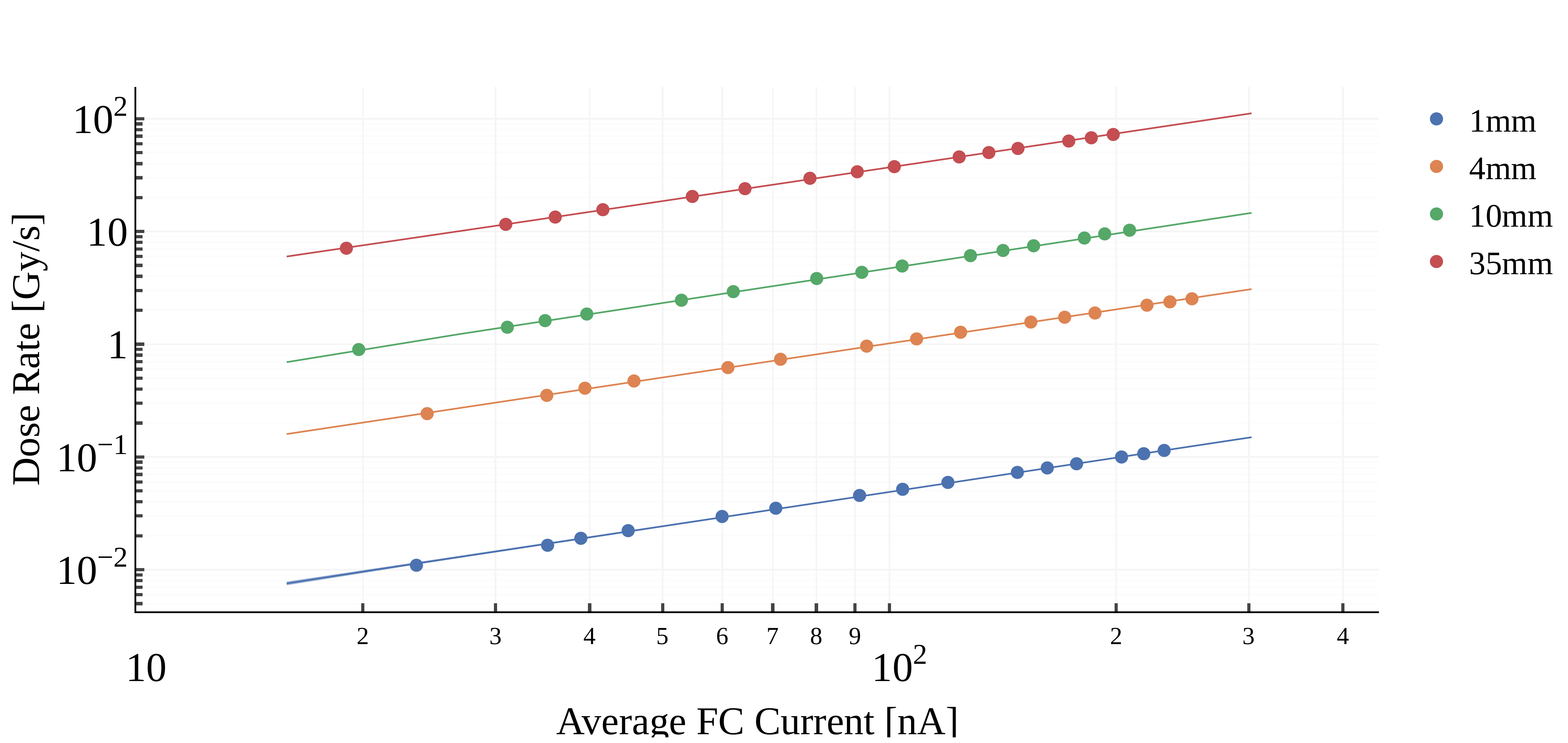}
\caption{Dose rate at the target position as a function of beam current measured upstream of the collimator, for collimator diameters of 1, 4, 10, and \SI{35}{\milli\meter}. Dose rate is measured by the IC, using Eq. \ref{eq:ICcal}, while the beam current is measured by the FC. Error bars are plotted but not visible. Lines represent linear fits, with the fit uncertainty shown as shaded regions.}
\label{fig:dose-rate-ranges}
\end{figure}

The resulting dose rate as a function of upstream beam current is shown in Figure~\ref{fig:dose-rate-ranges}. A linear dependence is observed over the full investigated range for all collimator apertures. 

These results demonstrate that, with the present passive scattering configuration, the dose rate can be continuously tuned from conventional radiotherapy levels (approximately 0.01~Gy/s) to well beyond the commonly cited FLASH threshold (approximately 40~Gy/s). 

This calibration can be used to monitor the average dose rate in the extracted field with an uncertainty of approximately 1\% using the in-beam IC. The subsequent translation of this value to estimate the dose delivered to a biological target is subject to uncertainty contributions from beam uniformity, target and diagnostics positioning accuracy, and LET-dependent corrections.

In this study, the maximum ion source current was limited to 29~mA to reduce activation of the collimator material. This represents an operational constraint rather than a technical limitation, and higher dose rates are achievable.

The minimum achievable dose rate could be further reduced by employing smaller collimator apertures or by increasing the scattering strength using a thicker or higher-Z foil. However, increased scattering would also reduce the proton energy and penetration depth, introducing a trade-off between dose-rate flexibility and beam energy.

\subsection{Gafchromic film quenching compensation} 

\begin{figure}[h]
\centering
\includegraphics[width=1.0\textwidth]{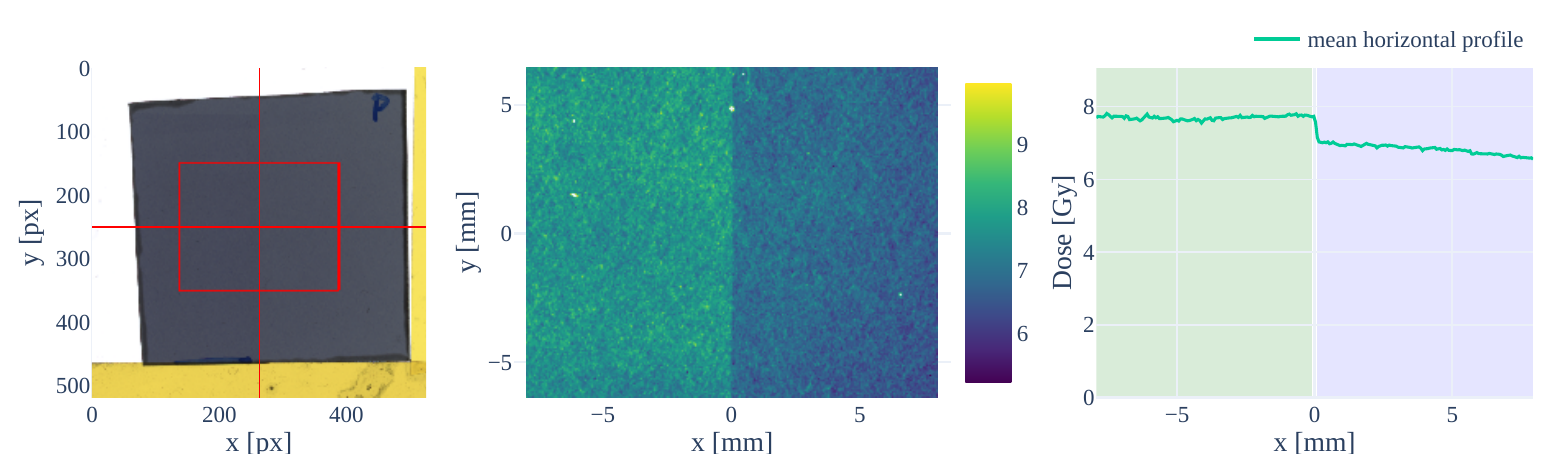}
\caption{Measurement of the lamination layer effect on film response. (left) Scanned film with rectangular region of interest (ROI) indicated. (centre) Corresponding dose map. (right) Mean horizontal dose profile; shaded regions correspond to laminated and non-laminated areas used to determine the dose ratio.}
\label{fig:lamination-results}
\end{figure}

The scanned film shows a clear increase in optical density in the region covered by the additional lamination layer, compared with the uncovered region. The scanned film and the corresponding dose profile are shown in Figure \ref{fig:lamination-results}.

From this profile, the dose ratio between the laminated and non-laminated sections of the film was determined to be 0.887(17). This value is consistent, within uncertainty, with the dose ratio between markers B and C expected from the LET-based simulations described in Section \ref{sec:gaf-methods}, supporting the interpretation that the observed dose difference arises from the increased LET after additional energy loss in the lamination material.


When the relative efficiency of 0.903(12) at 7.19 MeV (the proton energy incident on the active layer during a normal in-vitro irradiation) is also taken into account \citep{SanchezParcerisa2021}, the overall correction factor for the film response becomes 1.08(6). This factor must be applied to the measured film dose to obtain the dose delivered to the in-vitro target in this configuration. The correction factor is specific to the beam energy and material configuration used in this study, and must be recalculated if the beam or target geometry is changed.

Taking into account the uniformity of the field (Section \ref{sec:uniformity-results}), the overall uncertainty in the dose delivered to cells in a flask in this configuration is 9.46\%.

\subsection{SFRT tests}
\begin{figure}[h]
    \centering
    \begin{subfigure}[b]{0.25\textwidth}
           \includegraphics[width=\textwidth]{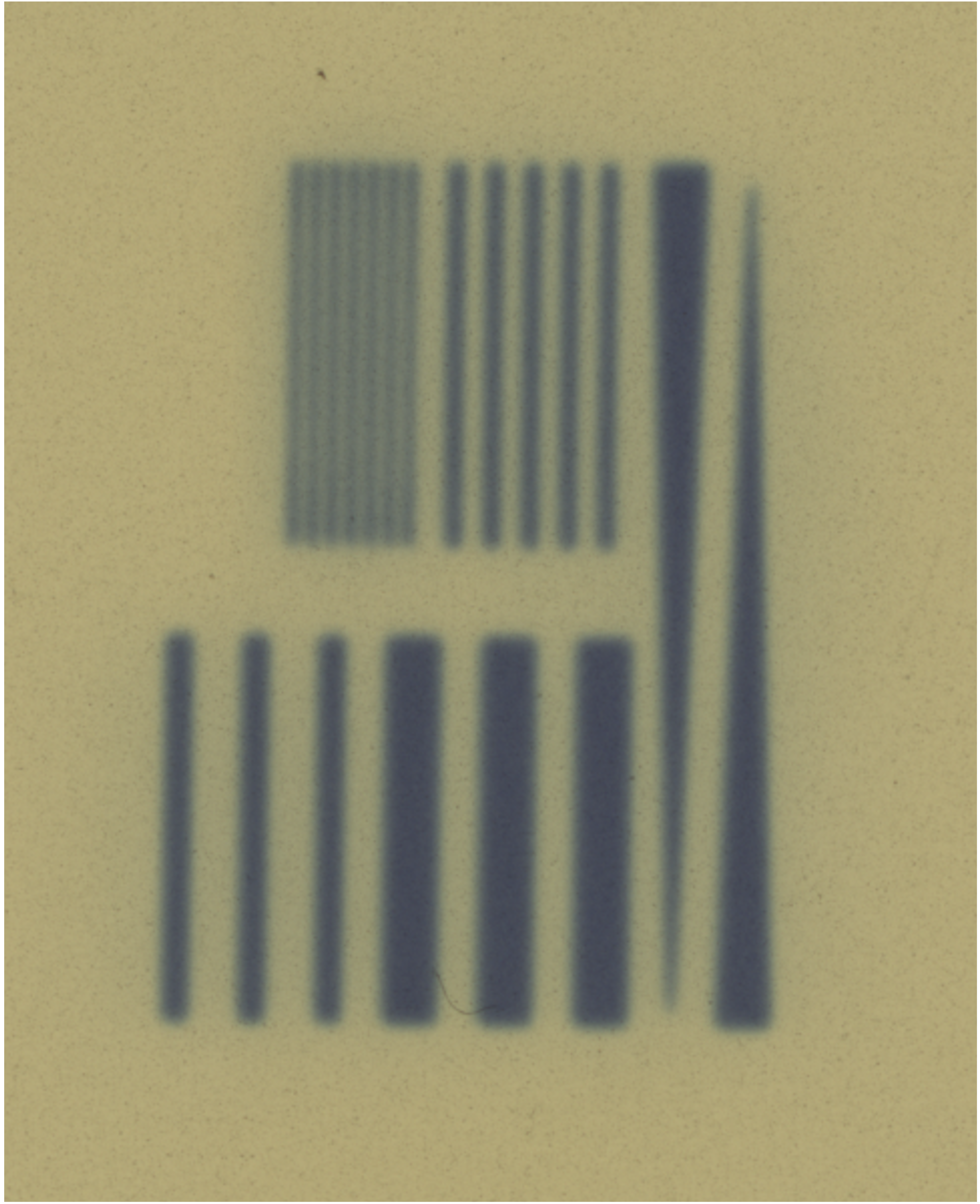}
        \caption{}
        \label{fig:subfig:sfrt-scan}
    \end{subfigure}
    \hfill
    \begin{subfigure}[b]{0.7\textwidth}
        \includegraphics[width=\textwidth]{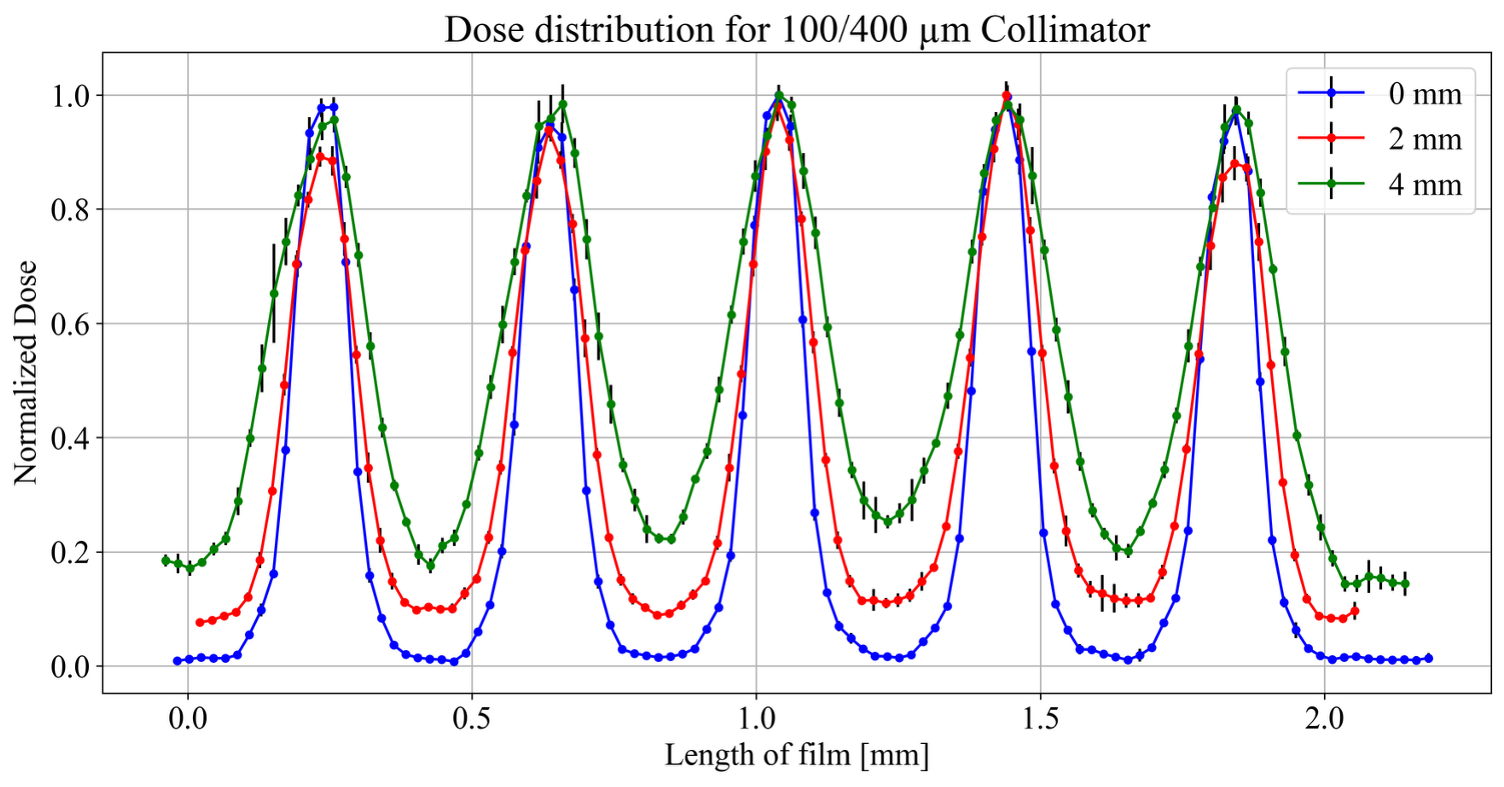}
        \caption{}
        \label{fig:subfig:sfrt-profile}
    \end{subfigure}
    \caption{SFRT beam structures obtained with the test collimator. (a) Film irradiated at a collimator-to-target distance of 4 mm; aperture widths and centre-to-centre spacings (\SI{}{\micro\meter}) are indicated. (b) Dose profiles for the 100/\SI{400}{\micro\meter} configuration at collimator-to-target distances of 0, 2, and \SI{4}{\milli\meter}. 
   \label{fig:sfrt-results}}
\end{figure}

Figure \ref{fig:sfrt-results} presents measurements of spatially fractionated radiation therapy (SFRT) beam structures at the BMC. An EBT-3 film was irradiated through the prototype collimator shown in Figure \ref{fig:sfrt-coll}, with the film positioned \SI{4}{\milli\meter} downstream of the collimator. The resulting film scan (Figure \ref{fig:subfig:sfrt-scan}) shows that all investigated grid patterns remain clearly resolved at this distance. This demonstrates that SFRT delivery is feasible under the investigated geometric conditions, even when the collimator cannot be placed directly in contact with the target, as would be the case in realistic in-vitro or in-vivo experimental geometries.

Although peaks and valleys remain spatially distinguishable at all distances, the peak-to-valley dose ratio (PVDR) is highly sensitive to the collimator–target separation. Figure \ref{fig:subfig:sfrt-profile} shows dose profiles for the 100/\SI{400}{\micro\meter} grid at separations of 0, 2, and \SI{4}{\milli\meter}. With increasing distance, a progressive rise in relative valley dose is observed due to lateral proton scattering in air. This effect is less dramatic for broader collimator spacings such as the 500/\SI{1000}{\micro\meter} as there is less overlap of the lateral penumbra of adjacent dose peaks. 

\begin{figure}[h]
    \centering
    \begin{subfigure}[h]{0.7\textwidth}
           \includegraphics[width=\textwidth]{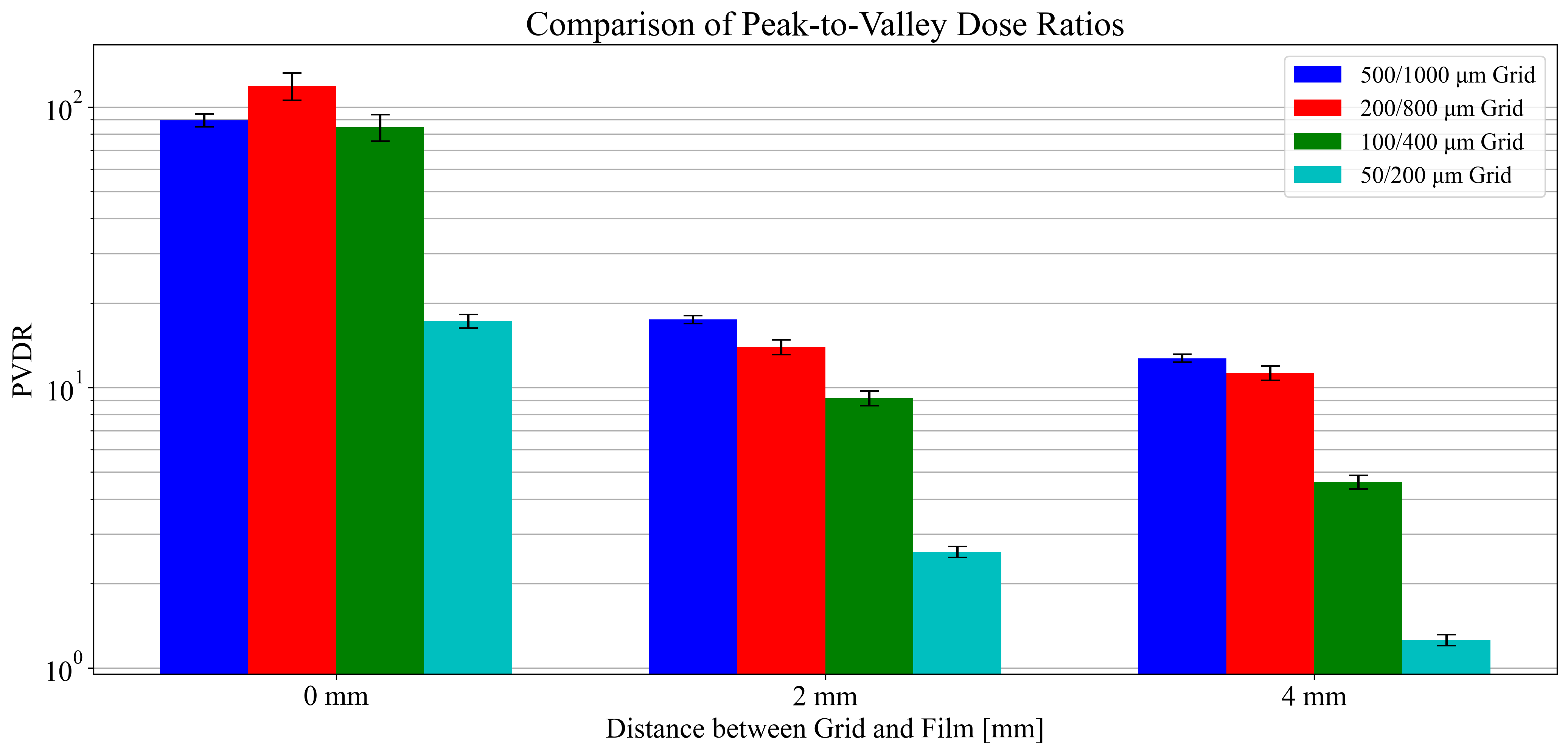}
        \caption{}
        \label{fig:subfig:pvdr-summary}
    \end{subfigure}
    \vfill
    \begin{subfigure}[h]{0.7\textwidth}
        \includegraphics[width=\textwidth]{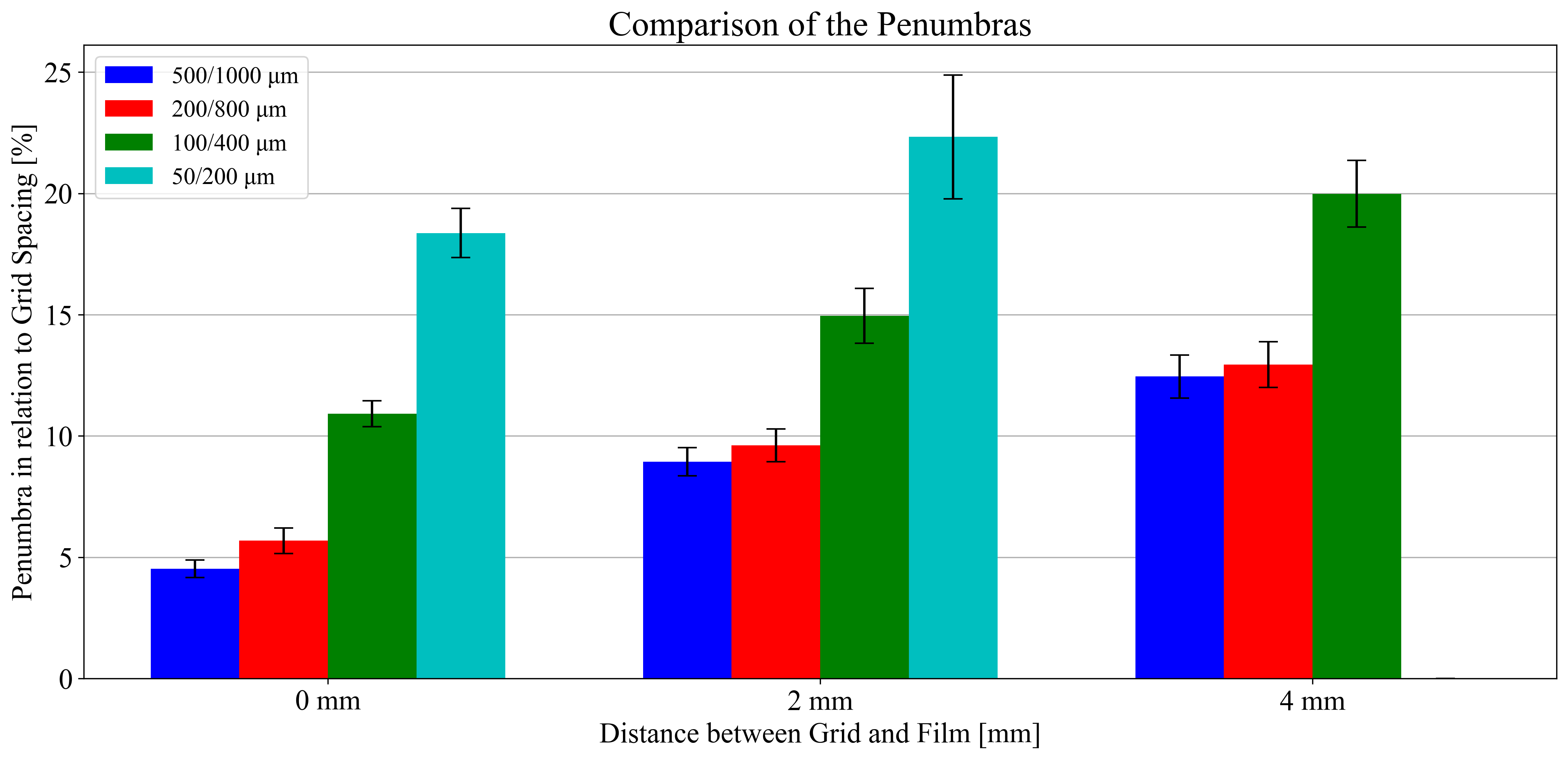}
        \caption{}
        \label{fig:subfig:penumbra-summary}
    \end{subfigure}
    \caption{Figure 12: Quantification of SFRT beam characteristics from EBT-3 film measurements. (a) Peak-to-valley dose ratio (PVDR) as a function of collimator-to-target distance for different grid spacings. (b) Lateral penumbra width (80\%–20\% of peak dose) normalised to the centre-to-centre spacing. No data are shown for the 50/\SI{200}{\micro\meter} grid at \SI{4}{\milli\meter} due to insufficient peak–valley separation. 
   \label{fig:sfrt-summary}}
\end{figure}

Figure \ref{fig:subfig:pvdr-summary} summarizes the experimentally measured PVDR for each grid spacing as a function of the collimator-to-target distance. The peak dose and valley dose are defined as the local maxima and minima, respectively, averaged over all peaks in the grid. The reduction of PVDR with increasing distance, due to beam divergence, is more pronounced for finer grid spacings.

This can also be visualized in Figure \ref{fig:subfig:penumbra-summary}, which quantifies the effect of beam divergence on the lateral penumbra of the dose peaks. The width of the penumbra is defined as the average distance over which the dose decreases from 80\% to 20\% of the peak dose, and is normalised to the centre-to-centre distance of each grid spacing. Once again, we see a clear increase in the relative penumbra as a function of distance, with the effect being more severe for finer grid spacings. For the 50/\SI{200}{\micro\meter} grid, the penumbra could not be quantified because the adjacent peaks were not fully resolved. 

These results indicate that SFRT implementations require careful control of the distance between the collimator and the irradiation target. To ensure reproducible PVDR values, this spacing should be minimized and mechanically fixed with priority given to positioning reproducibility. Without such control, small geometric variations may introduce systematic differences in valley dose and therefore in the biological outcome.

\section{Discussion}
\subsection{Beam shaping}
The uniformity of the extracted beam across the central region is approximately 10\%. This can be improved by increasing the drift space available for the scattered beam or by using a higher-Z scattering material. Both approaches enhance lateral beam spread, reducing fluence gradients across the field.  In addition, improved scattering efficiency allows a thinner scatterer to be used while maintaining adequate uniformity, which in turn permits extraction at higher proton energies.

Operation at higher proton energies is advantageous for several reasons. The stopping power gradient increases non-linearly as proton energy decreases, making dose deposition highly sensitive to small energy uncertainties near the Bragg peak. Extracting at higher energies therefore reduces depth–dose sensitivity to energy fluctuations and improves dosimetric robustness. Furthermore, higher energies enable deeper penetration, allowing irradiation of thicker or more complex targets.

An upgraded beam shaping and monitoring unit is currently under development to address these limitations and to streamline the irradiation workflow. The system will include interchangeable collimators and scatterers, temperature monitoring, and real-time proton current readout. Remote control of these components is expected to improve reproducibility, reduce setup time, and enable more consistent beam conditions, reducing the number of required calibration measurements.

\subsection{Dosimetry}
The primary challenge in this proof-of-concept setup was achieving accurate and reliable dosimetry. Many established proton beam characterization techniques are optimized for higher energies than those available at the BMC laboratory. At the lower energies used here, the shallow penetration depth and elevated LET complicate detector response and calibration. Moreover, detectors suitable for conventional dose-rate measurements may experience saturation under FLASH conditions, further limiting their applicability across operating modes.

This limitation is particularly critical in radiobiological experiments, where dose–response relationships can be steep and small systematic dose errors may lead to significant misinterpretation of biological effects. At low doses, a 9.46\% uncertainty would have limited impact—for example, 2 Gy would be delivered with a 0.2 Gy error. At high doses such as 10 Gy, however, the corresponding 1 Gy error could produce significant biological differences in the measured endpoints. Work is already underway to reduce this uncertainty and improve the reliability of biological readouts in future experiments.

The Gafchromic EBT-3 films used in this work were calibrated using X-rays. However, their response to low-energy protons is affected by significant quenching relative to photon irradiation. As shown in Section \ref{sec:gaf-methods}, the film response is also influenced by energy loss in overlying materials, which modifies the LET within the active layer. Both effects introduce energy-dependent correction factors that must be applied to convert film response to absorbed dose. For the present configuration, the combined correction factor was determined to be 1.08(6), but this value is valid only for a beam energy of approximately \SI{8.14}{\mega\electronvolt} at the target and the specific material stack used.

These correction factors are specific to the proton energy spectrum and material configuration of the experimental setup. Changes in beamline elements, air gaps, detector positions, or target holders will alter the energy at the point of irradiation and therefore require renewed characterization. For this reason, systematic measurements of film response as a function of proton energy for this beamline are planned over the range of available energies (approximately 3-18 MeV) once the final experimental configurations for in-vitro and in-vivo studies are established.

Future improvements will address dosimetric uncertainties due to high LET by minimizing energy degradation in the setup where possible and by developing efficient energy verification protocols. In addition, efforts will be made to improve the reproducibility of the setup. For example, fixing the position of the monitoring IC in the beam path will reduce the need for recalibration. 

\subsection{SFRT}
 The beam profiles obtained with the 500/1000 and 200/800 grid spacings are significantly more robust to small geometric variations and should be used instead of the finer grid spacing options whenever possible to ensure reproducibility. In in-vivo irradiations, target geometry (tumour size, animal size, etc.) can easily exhibit variations on the order of 1~mm. The grid spacing should therefore be chosen to maximize robustness while ensuring sufficient peak coverage of the tumour volume to preserve the effects of spatial fractionation. 

\subsection{Clinical Translation}
The primary purpose of our facility is not to reproduce clinical treatment conditions, but to provide a controlled environment for systematically investigating the biological mechanisms underlying proton therapy. This is particularly important for SFRT, where the range of treatment parameters is substantially more complex than in conventional radiotherapy, with numerous dosimetric variables that may influence biological response. Clinical proton therapy facilities are generally less suited to such studies, as access to the Bragg peak typically requires extensive energy degradation, resulting in a broad energy spectrum and reduced experimental control. In addition, research activities must not interfere with routine patient treatments or clinical beamline operation, limiting the flexibility required for systematic experimental investigations. Systematic evaluation of these parameters is essential to establish a robust biological basis for treatment design. The resulting evidence provides a rational basis for selecting and justifying treatment parameters during clinical translation and can help guide the development of future clinical protocols.

\section{Conclusions and Outlook}
In this work, we have demonstrated the development of a flexible research platform for systematic in-vitro and in-vivo proton therapy studies at the BMC laboratory. Through passive beam shaping, interchangeable collimation, and a custom beam chopper system, the facility enables proton irradiations spanning conventional dose rates and ultra-high (FLASH) dose rates, as well as spatially-fractionated beam structures.


Comparative beam profile measurements show that passive scattering significantly improves uniformity and reduces sensitivity to extraction drifts relative to magnetic defocusing. Using the passive scattering configuration, the standard deviation of the delivered dose within a 20~mm radius of the beam spot center is evaluated at 7.96\% of the mean. This value can be further improved without reducing the extracted beam energy by selecting a higher-Z scattering material. An improved setup using a thin Tantalum foil for scattering is currently under development.

A key focus of this study was the development of a preliminary dosimetric framework for this non-standard energy range. Calibration of the in-beam ionization chamber enabled direct monitoring of the dose rate at the target, and systematic characterization across collimator apertures demonstrates continuous dose-rate tunability over 5 orders of magnitude. Because irradiation occurs near the Bragg peak, radiochromic film dosimetry requires careful correction for LET quenching and energy loss in overlying materials. The experimentally validated correction factor allows quantitative dose assessment for in-vitro configurations, though it remains geometry- and energy-specific. Its systematic characterization is essential for quantitative dosimetry. First in-vitro irradiations have already been performed to validate this preliminary dosimetry framework, and will be reported in a forthcoming publication. 

SFRT tests confirmed that well-resolved proton minibeam structures can be delivered at biologically relevant target distances, while highlighting the strong dependence of PVDR on collimator–target spacing. Together, these results establish the BMC beamline as a platform for controlled studies of proton radiobiology under varied spatial and temporal delivery conditions.

Ongoing work will focus on improved beam shaping and monitoring, improved beam energy measurement, reduction of dosimetric uncertainties through systematic characterization of reproducible experimental configurations, and expansion to in-vivo irradiation setups. Due to its flexibility and accessibility, this facility provides a controlled framework to foster advances in systematic in-vivo and in-vitro proton therapy studies, including investigations of the interplay between LET, dose rate, and spatial fractionation in proton radiobiology. 

\section{Acknowledgements}
We acknowledge contributions from LHEP engineering and technical staff. This research project was partially funded by the Swiss National Science Foundation (SNSF) (grant: IZURZ2\_224901) and by the BIND Grant programme of the University of Bern.\\\\


\medskip
\addcontentsline{toc}{section}{\numberline{}References}
\vspace*{-10mm}

\bibliographystyle{JHEP}
\bibliography{mappBib-2.bib}

\end{document}